\documentclass[11pt,letterpaper,twoside]{article}
\usepackage[centertags]{amsmath}
\usepackage{graphicx,indentfirst,amsmath,amsfonts,amssymb,amsthm,newlfont}
\font\Goth=yinitas scaled \magstep0

\newcommand{\Gth}[1]{\lower2mm\hbox{\Goth #1}}

\usepackage[latin1]{inputenc}
\def\al{\alpha}

\def\be{\beta}

\def\l1{{\lambda}_1}

\newcommand{\f}{\frac}

\def\x1{{\xi }_{xx}}
\def\x2{{\xi }_{yy}}
\def\x3{{\xi }_{xy}}

\def\e1{{\eta }_{xx}}
\def\e2{{\eta }_{yy}}
\def\e3{{\eta }_{xy}}

\newcommand{\ds}{\displaystyle }
\newtheorem{theorem}{Theorem}

\newtheorem{corollary}{Corollary}
\newcommand{\beqn}{\begin{eqnarray*}}
\newcommand{\eeqn}{\end{eqnarray*}}
\newcommand{\beqnn}{\begin{eqnarray}}
\newcommand{\eeqnn}{\end{eqnarray}}
\newcommand{\p}{\partial}
\newcommand{\bb}{\begin{equation}}
\newcommand{\ee}{\end{equation}}
\newcommand{\ba}{\begin{array}}
\newcommand{\ea}{\end{array}}
\newcommand{\R}{\mathbb{R}}

\begin{document}
\pagenumbering{arabic}
\title{\huge \bf Conservation laws for self-adjoint first order evolution equation}
\author{\rm \large Igor Leite Freire \\
\\
\it Centro de Matemática, Computação e Cognição\\ \it Universidade Federal do ABC - UFABC\\ \it 
Rua Catequese, $242$,
Jardim,
$09090-400$\\\it Santo André, SP - Brasil\\
\rm E-mail: igor.freire@ufabc.edu.br\\
}
\date{\ }
\maketitle
\vspace{1cm}
\begin{abstract}
In this work we consider the problem on group classification and conservation laws of the general first order evolution equations. We obtain the subclasses of these general equations which are quasi-self-adjoint and self-adjoint. By using the recent new conservation theorem due to Ibragimov, we establish the conservation laws of the equations admiting self-adjoint equations. We illustrate our results applying them to the inviscid Burgers' equation. In particular an infinite number of new symmetries of these equations are found and their corresponding conservation laws are established. \end{abstract}
\vskip 1cm
\begin{center}
{2000 AMS Mathematics Classification numbers:\vspace{0.2cm}\\
76M60, 58J70, 70G65\vspace{0.2cm} }
\end{center}
\pagenumbering{arabic}
\newpage

\section{Introduction}

\

The Lie point symmetries of evolution equations with one spatial variable
\bb\label{a1}
u_{t}=F\left(t,x,u,\f{\p u}{\p x},\cdots, \f{\p^{n}u}{\p x^{n}}\right),
\ee
with $n\geq 2$, has been studied by many authors, see \cite{blu, bruzon, cher,igor3,tor, gai, ib0, ib,la,iva,  qu, zh2, zh3} and references therein.

For example, if $n=2$, equation (\ref{a1}) includes the nonlinear heat equation, the Burgers equation, Fokker-Planck equation, Black-Scholes equation and, more generally, reaction-diffusion-convection equations, see \cite{cher, tor, ib, iva, zh2, zh3}.

The Korteweg-de Vries (KdV) equation, cylindrical KdV and modified KdV are examples of third order evolution equations, see \cite{blu, ib}.

When $n=4$, equation (\ref{a1}) includes the modified Kuramoto-Sivashinsky equation, the Cahn-Hilliard equation, the thin film equation and others, see \cite{ bruzon, gai, qu}.

However, the first order equation
\bb\label{1.1}
u_{t}+f(t,x,u,u_{x})=0
\ee
seems to have received few attention.

To the best of our knowledge, the earliest work involving first order evolution equations and Lie point symmetries was \cite{ou}, where the authors studied equation (\ref{1.1}) with $f=a(u)u_{x}$, for some particular functions $a(u)$. After it, Nadjafikhah \cite{me1} obtained projectable symmetries of equation
\bb\label{3}
u_{t}+a(u)u_{x}=0
\ee
and in \cite{me2} the same author classifies the similarity solutions of the symmetries obtained in \cite{me1}. Equation (\ref{3}) is known as inviscid Burgers equation.

The purpose of this work is to deal with the problem on group classification of the general first order evolution equation and how to obtain conservation laws from the Lie point symmetries. We intend to
\begin{itemize}

\item find the first order evolution equations that admit (quasi) self-adjoint equations;

\item obtain close formulae to express conservation laws for equations of the type (\ref{1.1}) using recent results due to Ibragimov \cite{ib};

\item generalize the previous results on group classification of equation (\ref{3});

\item establish conservation laws for equation (\ref{3}).
\end{itemize}

The paper is organized as the follows. In the next section we obtain the general determining equations for the components of symmetry generators, the (quasi) self-adjointness condition and establish the corresponding conservation laws for self-adjoint equations of the type (\ref{1.1}). We also obtain new Lie point symmetry generators of (\ref{3}) and some conservation laws for it are established.

\section{Main results}\label{main}

\

In this section we shall consider the group classification problem and how to find conservation laws for equation (\ref{1.1}) with $f_{u_{x}}\neq 0$. Hereafter all functions will be assumed to be smooth, the summation over the repeated indices is understood, $u_{x}=\f{\p u}{\p x}$ and $u_{t}=\f{\p u}{\p t}$.

Following the standard Lie approach \cite{bk,ibt,ol}, let
\bb\label{2.2}
X=\tau(t,x,u)\f{\p}{\p t}+\xi(t,x,u)\f{\p}{\p x}+\eta(t,x,u)\f{\p}{\p u}
\ee
be a Lie point symmetry generator of equation (\ref{1.1}). Then the coeficients $\tau,\,\xi$ and $\eta$ satisfy the following equation
\bb\label{2.3}
\ba{l}
\eta_{t}-\xi_{t}u_{x}+(\tau_{t}-\eta_{u}+\xi_{u}u_{x})f+\xi f_{x}+\tau f_{t}+\eta f_{u}-\tau_{u}f^{2}\\
\\
+(\eta_{x}+\eta_{u}u_{x}-\xi_{x}u_{x}-\xi_{u}u_{x}^{2})f_{u_{x}}+(\tau_{x}+\tau_{u}u_{x})ff_{u_{x}}=0.
\ea
\ee

We observe that (\ref{2.3}) is one equation to be solved for $4$ unknown functions $\xi,\,\tau$, $\eta$ and $f$.

To obtain the full symmetry group of equation (\ref{1.1}) it is necessary to find all possible functions $\xi,\,\tau,\,\eta,\,f$ satisfying the relation (\ref{2.3}). Thus, a complete group classification of (\ref{1.1}) is impossible.

Let us now consider the problem of finding conservation laws for equations of the type (\ref{1.1}).

If an equation possesses variational structure, the Noether Theorem can be employed in order to establish conservation laws for it, {\it e.g.}, see \cite{igor1, yi2,igor2, naz}.

However, it is well-known that evolution equations do not possess variational structure. Then, they cannot be obtained from the Euler-Lagrange equations and the Noether's Theorem cannot be applied to them in order to obtain conservation laws.

Fortunately there are some alternative methods to obtain conservation laws for equations without Lagrangians: the direct method, the characteristic method, the variational approach, the symmetry conditions, the direct construction formula and the partial Noether approach. For a more detailed discussion, see \cite{blu, ib, naz}. Although some of these methods could be employed in order to establish conservation laws for equation (\ref{3}), in this paper we shall use recent results due to Ibragimov \cite{ib} in order to construct conservation laws for equations of the type (\ref{1.1}). We shall refer to the new conservation theorem established in \cite{ib} (see Theorem 3.5 in the reference) as Ibragimov's Theorem.

The Ibragimov's Theorem on conservation laws can be summarized by the following algorithm (see \cite{ib} for more details): given a PDE $$F=F(x,u,\p u,\cdots, \p^{n} u)=0,$$ where $\p^{k} u$ denotes the set of all $kth$ order derivatives of $u$,
\begin{itemize}
\item we construct a Lagrangian ${\cal L}=vF$.
\item From the Euler-Lagrange equations, the following system is obtained:
\bb\label{eq00}
F(x,u,\p u,\cdots, \p^{n} u)=0,
\ee
\bb\label{eq01}
F^{\ast}(x,u,v,\p u,\p v,\cdots, \p^{n} u,\p^{n}v)=0.
\ee

The second equation of the system (\ref{eq00}) - (\ref{eq01}) is called adjoint equation to $F=0$.

Equation (\ref{eq01}) is said to be quasi-self-adjoint if the system (\ref{eq00}) - (\ref{eq01}) is equivalent to the original equation (\ref{eq00}) upon the substitution $v=\varphi(u)$ such that $\varphi'(u)\neq 0$, {\it i.e},
\bb\label{cond}
\left.F^{\ast}(x,u,v,\p u,\p v,\cdots, \p^{n} u,\p^{n}v)\right|_{v=\varphi(u)}=\phi F(x,u,\p u,\cdots, \p^{n} u),
\ee
for some function $\phi=\phi(x,u,\p u,\cdots,\p^{n} u)$.

If equation (\ref{cond}) is true with $\varphi(u)=u$, then (\ref{eq00}) is said to be self-adjoint. For more details, see \cite{bruzon, ib, ib1}.

\item The conserved vector is $C=(C^{i})$, where
\bb\label{2.4}
\ba{l c l}
C^{i}&=&\ds{\xi^{i}{\cal L}+W\left[\f{\p{\cal L}}{\p u_{i}}-D_{j}\left(\f{\p{\cal L}}{\p u_{ij}}\right)+D_{j}D_{k}\left(\f{\p{\cal L}}{\p u_{ijk}}\right)-\cdots\right]}\\
\\
&&
\ds{+D_{j}(W)\left[\f{\p{\cal L}}{\p u_{ij}}-D_{k}\left(\f{\p{\cal L}}{\p u_{ijk}}\right)+\cdots\right]+\cdots}
\ea
\ee
and $W=\eta-\xi^{j}u_{j}$.
\end{itemize}

Then, applying this algorithm to equation (\ref{1.1}), we obtain:
\begin{itemize}
\item Lagrangian: \bb\label{lag}{\cal L}=vu_{t}+vf(t,x,u,u_{x}).\ee
\item Adjoint equation: the adjoint equation to (\ref{1.1}) is $F^{\ast}=0$, where
\bb\label{adj}
F^{\ast}=-v_{t}-v_{x}f_{u_{x}}+vf_{u}-vf_{xu_{x}}-vu_{x}f_{uu_{x}}-vf_{u_{x}u_{x}}u_{xx}.
\ee
\item Components of the conserved vector $C=(C^{0},C^{1})$:
\bb\label{2.5}
\ba{l c l}
C^{0}&=&(\eta+\tau f-\xi u_{x})v,\\
\\
C^{1}&=&(\eta+\tau f-\xi u_{x})vf_{u_{x}}.
\ea
\ee
\end{itemize}

\subsection{Quasi-self-adjointness condition of equation $(\ref{1.1})$}

Supposing that $$\left.F^{\ast}\right|_{v=\varphi(u)}=\phi F,$$
where $F=u_{t}+f(t,x,u,u_{x})$ and $F^{\ast}$ is given by (\ref{adj}), we obtain $\phi=-\varphi'(u)$ and
\bb\label{2.2.1}
\left\{\ba{l}
f_{u_{x}u_{x}}=0,\\
\\
\varphi(u)f_{u}-\varphi'(u) f_{u_{x}}u_{x}-\varphi(u)f_{uu_{x}}u_{x}-\varphi(u)f_{xu_{x}}+\varphi'(u)f=0.
\ea\right.
\ee

From (\ref{2.2.1}) we conclude that $f=\al(t,x,u)u_{x}+\be(t,x,u)$, with $\al\neq0$. Hence, the functions $\varphi(u),\,\al(t,x,u)$ and $\be(t,x,u)$ should satisfy
\bb\label{2.2.2.1}
\be_{u}\varphi(u)+\varphi'(u)\be=\varphi(u)\al_{x}.
\ee

It follows that if (\ref{1.1}) is quasi-self-adjoint, we have two cases to consider:

{\bf Case 1}: If $\be\neq 0$, in order for (\ref{2.2.2.1}) to be true, then
\bb\label{*}
\f{\al_{x}-\be_{u}}{\be}=\f{\varphi'(u)}{\varphi(u)},
\ee
and in this case
\bb\label{eq2.0.0}
\varphi(u)=c\,\exp{\int\f{\al_{x}-\be_{u}}{\be}\,d u},
\ee
where $c\in\R$ is an arbritrary constant.

{\bf Case 2}: If $\be=0$, from (\ref{2.2.2.1}), $\al=\al(t,u)$ and $\varphi$ is an arbitrary function.

Reciprocally, if $f=\al(t,u)u_{x}$ in (\ref{1.1}), it is easy to check that (\ref{1.1}) is quasi-self-adjoint. When $f=\al(t,x,u)u_{x}+\be(t,x,u)$, then $f$ satisfies (\ref{2.2.1}) if (\ref{*}) is satisfied and then, function $\varphi$ is given by (\ref{eq2.0.0}).

According to equation (\ref{2.5}), taking $v=\varphi(u)$ (quasi-self-adjointness condition), a conservation law for equation
\bb\label{2.2.2}
u_{t}+\al(t,x,u)u_{x}+\be(t,x,u)=0,
\ee
is $D_{t}C^{0}+D_{x}C^{1}=0$, where $\al$ and $\be$ are as considered in cases 1 or 2, and
\bb\label{2.3.1}
\ba{l c l}
C^{0}&=&[\eta+\tau \be+(\tau\al-\xi)\, u_{x}]\,\varphi(u),\\
\\
C^{1}&=&[\eta\al+\xi \be-(\tau\al-\xi)\, u_{t}]\,\varphi(u).
\ea
\ee

\subsection{Self-adjointness condition of equation $(\ref{1.1})$}

Let us now find the class of the self-adjoint equations of the type (\ref{1.1}). Whenever $\varphi=u$, equation (\ref{2.2.2.1}) becomes
$$\be_{u}\,u+\be=u\,\al_{x}.$$

Again we have two cases to consider:

{\bf Case 1}: If $\be\neq 0$, then
\bb\label{**}
\be=\f{1}{u}\int u\,\al_{x}d u+\f{\lambda(t,x)}{u},
\ee
for some function $\lambda=\lambda(t,x)$.

{\bf Case 2}: If $\be=0$ then $\al=\al(t,u)$.

From equation (\ref{**}), we observe that the case $\be=0$ occurs if and only if $\al_{x}=\lambda=0$. Then, the most general form of a self-adjoint equation of the type (\ref{1.1}) is (\ref{2.2.2}), where $\be$ is given by (\ref{**}).

It is easy to check that all equations of the type (\ref{2.2.2}), with $\be$ satisfying (\ref{**}), are self-adjoints.

Equation (\ref{2.2.2}) includes
\begin{itemize}
\item inviscid Burgers equation, taking $\al=a(u)$ and $\be=0$, see \cite{ ou, me1, me2,fabio1};
\item linear transport equation, taking $\al=q(x)$ and $\be=0$, see \cite{fabio2, fabio3}.
\end{itemize}

From  equation (\ref{2.3.1}), taking $\varphi=u$ (self-adjointness condition), a conservation law for equation (\ref{2.2.2}) is $D_{t}C^{0}+D_{x}C^{1}=0$, where
\bb\label{2.3.1'}
\ba{l c l}
C^{0}&=&[\eta+\tau \be+(\tau\al-\xi)u_{x}]\, u ,\\
\\
C^{1}&=&[\eta\al+\xi \be-(\tau\al-\xi) u_{t}]\,u.\ea
\ee

\subsection{Theorems on self-adjoint equations and conservation laws}

Our main results on the self-adjointness conditions and conservation laws can be summarized by the following theorems.

\begin{theorem}\label{teo1}
Let $(\ref{2.2})$ be a Lie point symmetry generator of equation $(\ref{1.1})$. Then the symmetry coefficients satisfy $(\ref{2.3})$.
\end{theorem}

\begin{corollary}\label{cor1}
The determining equations of $(\ref{2.2.2})$ are given by
\bb\label{deteq}
\ba{l}
\eta_{t}+\be(\tau_{t}-\eta_{u})+\be_{x}\,\xi+\be_{t}\,\tau+\be_{u}\,\eta-\be^{2}\tau_{u}+\al\,\eta_{x}+\al\,\be\,\tau_{x}=0,\\
\\
-\xi_{t}+\al\,\tau_{t}+\be\,\xi_{u}+\al_{x}\,\xi+\al_{t}\,\tau+\al_{u}\,\eta-\al\,\be\,\tau_{u}-\al\,\xi_{x}+\al^{2}\,\tau_{x}=0.
\ea
\ee
\end{corollary}
{\bf Proof} Substituting $f=\al(t,x,u)u_{x}+\be(t,x,u)$ into (\ref{2.3}), we obtain (\ref{deteq}). $\square$

\begin{theorem}\label{teo1'}
The following statements about equation $(\ref{2.2.2})$ are true:
\begin{enumerate}
\item If $\be=0$, $(\ref{2.2.2})$ is quasi-self-adjoint if and only if $\al=\al(t,u)$.
\item If $\be\neq 0$, $(\ref{2.2.2})$ is quasi-self-adjoint if and only if the functions $\al$ and $\be$ satisfy the relation $(\ref{*})$, for some function $h=h(u)$, and $\varphi$ is given by $(\ref{eq2.0.0})$.
\end{enumerate}
\end{theorem}

\begin{theorem}\label{teo2}
Equation $(\ref{1.1})$ is self-adjoint if and only if $f=\al(t,u) u_{x}$ or $f=\al u_{x}+\be$, where $\be$ is given by $(\ref{**})$.
\end{theorem}

\begin{theorem}\label{teo4}
A conservation law for the system
$$
\left\{
\ba{l}
u_{t}+f(t,x,u,u_{x})=0,\\
\\
-v_{t}-v_{x}f_{u_{x}}+vf_{u}-vf_{xu_{x}}-vu_{x}f_{uu_{x}}-vf_{u_{x}u_{x}}u_{xx}=0.
\ea
\right.
$$
is $Div(C)=D_{t}C^{0}+D_{x}C^{1}=0$, where $C^{0}$ and $C^{1}$ are given by $(\ref{2.5})$ and $\tau, \,\xi$ and $\eta$ are the coefficients of the generator $(\ref{2.2})$.
\end{theorem}

\begin{theorem}\label{teo5}
A conservation law for equation $(\ref{2.2.2})$, with $\al$ and $\be$ as in Theorem $\ref{teo2}$, is $Div(C)=D_{t}C^{0}+D_{x}C^{1}=0$, where $C^{0}$ and $C^{1}$ are given by $(\ref{2.3.1'})$ and $\tau, \,\xi$ and $\eta$ are the coefficients of the generator $(\ref{2.2})$.
\end{theorem}

\subsection{Inviscid Burgers equation}

We recall that to obtain the group classification of (\ref{1.1}) we need to construct all possible functions $\xi,\tau,\,\eta$ and $f$ obeying (\ref{2.3}). As mentioned above, this is an underdetermined problem and a general group classification is impossible.

Regarding equation (\ref{2.2.2}), from Corollary \ref{cor1}, we conclude that it possesses an infinity dimensional symmetry Lie algebra.

Equation (\ref{2.2.2}) covers the so-called inviscid Burgers equation and it describes turbulence phenomena, for instance, compressible gas dynamics, shallow water flow, weather prediction, plasma modelling, rarefied gas dynamics and many others, see \cite{fabio1, fabio2, fabio3, me1, me2,  ned, sar1, sar2, shen, zah}.

In \cite{fabio1} the random Riemann problem for Burgers equation is solved. In \cite{fabio2} a numerical scheme to approximate the $m$th moment of the solution of the one-dimensional random linear transport equation is studied. In \cite{fabio3} a numerical scheme for the random linear transport equation is presented. In \cite{zah} numerical methods are employed for solving hyperbolic conservation laws.

Distributional products and solutions of the inviscid Burgers equation
\bb\label{b1}
u_{t}+uu_{x}=0
\ee
are studied in \cite{sar1, sar2}. In \cite{sar1} the concept of global $\al$-solution for equation (\ref{b1}) is introduced, as well as the existence of ``delta-soliton'' travelling waves. In \cite{sar2} new solutions are presented. System of transport equations are considered in \cite{ned,shen}. Further details can be found in the references cited above. In what follows, the Lie point symmetries and conservation laws for (\ref{3}) shall be discussed.

\subsubsection{Projectable symmetries of inviscid Burgers equation}

Let us now consider the symmetries of equation (\ref{3}).

From equation (\ref{2.3}), the symmetry coefficients $\tau,\,\xi$ and $\eta$ satisfy the following determining equations
\bb\label{3.2}
\eta_{t}+a(u)\eta_{x}=0,
\ee
\bb\label{3.3}
\eta a'(u)+\tau_{x}a(u)^{2}-\xi_{t}+\tau_{t}a(u)-\xi_{x}a(u)=0.
\ee

Since the system (\ref{3.2}) - (\ref{3.3}) is underdetermined, we use the symmetries obtained by Nadjafikhah in \cite{me1}. The {\it ansatz} employed by Nadjafikhah in \cite{me1} was to consider the projectable symmetries of (\ref{3}). For more details, see \cite{me1}.

Supposing that $\tau=\tau(t,x)$ and $\xi=\xi(t,x)$, Nadjafikhah obtained the following basis to the symmetry Lie algebra (see \cite{me1,me2}):

\bb\label{s1}
\ba{l}
\ds{X_{1}=\f{\p}{\p t},\,\,\,X_{2}=\f{\p}{\p x},\,\,\,\,X_{3}=x\f{\p}{\p x}+t\f{\p }{\p t},\,\,\,\,X_{4}=t\f{\p}{\p t}-\f{a(u)}{a'(u)}\f{\p }{\p u},}\\
\\
\ds{X_{5}=t\f{\p}{\p x}+\f{1}{a'(u)}\f{\p }{\p u},\,\,\,\,X_{6}=x\f{\p}{\p t}-\f{a(u)^{2}}{a'(u)}\f{\p }{\p u},}\\
\\
\ds{X_{7}=t^{2}\f{\p}{\p t}+tx\f{\p}{\p x}+\f{x-ta(u)}{a'(u)}\f{\p }{\p u},}\\
\\
\ds{X_{8}=tx\f{\p}{\p t}+x^{2}\f{\p}{\p x}+\f{a(u)(x-ta(u))}{a'(u)}\f{\p }{\p u}.}
\ea
\ee

Two questions naturally arise:

{\bf Q1}: Are there symmetries such that $(\xi_{u},\tau_{u})\neq (0,0)$?

{\bf Q2}: Could it possible to find symmetries more general than that obtained in \cite{me1}?

An (simple) answer to {\bf Q1} is the following: suppose $\tau=\tau(u)$ and $\xi=\xi(u)$. From (\ref{3.2}) - (\ref{3.3}) we conclude that $\eta=0$ and
$$X=\tau(u)\f{\p}{\p t}+\xi(u)\f{\p}{\p x}$$
is a Lie point symmetry generator of (\ref{3}).

In the next subsection {\bf Q2} shall be answered.

\subsubsection{New Lie point symmetry generators for equation $(\ref{3})$}

With regard to {\bf Q2}, according to Corollary \ref{cor1}, if $\be=0$ in (\ref{2.2.2}) and
\bb\label{proj1}
X=\tau(t,x)\f{\p}{\p t}+\xi(t,x)\f{\p}{\p x}+\eta(t,x,u)\f{\p}{\p u}
\ee
is a projectable symmetry generator of (\ref{2.2.2}), then the determining equations (\ref{2.3}) do not have terms involving derivatives of the Lie point symmetry generators with respect to $u$. Thus it is easy to check that the components of the vector field
$$X_{\lambda}=\lambda(u)\,\tau(t,x)\f{\p}{\p t}+\lambda(u)\,\xi(t,x)\f{\p}{\p x}+\lambda(u)\,\eta(t,x,u)\f{\p}{\p u},$$
where $\lambda=\lambda (u)$ is a smooth function, satisfy the determinig equations (\ref{deteq}). So, the field $X_{\lambda}$ is a nonprojectable Lie point symmetry generator and the following results are immediate consequences of the Corollary \ref{cor1}, Theorem \ref{teo2} and Theorem \ref{teo5}.

\begin{theorem}\label{teo6}
Let $X$ be a projectable Lie point symmetry generator of equation
\bb\label{111}
u_{t}+\al(t,u)u_{x}=0,
\ee
and $\lambda(u)$ a smooth function. Then the vector field $X_{\lambda}=\lambda(u)X$ is a Lie point symmetry of the equation $(\ref{111})$.
\end{theorem}

\begin{corollary}\label{cor2}

Let $(\ref{proj1})$ be a projectable Lie point symmetry generator of equation $(\ref{111})$, $\lambda(u)$ a smooth function and
\bb\label{ncl}
\ba{lcl}
C^{0}&=&\lambda(u)[\eta+(\tau \al-\xi)\, u_{x}]\,u,\\
\\
C^{1}&=&\lambda(u)[\eta-(\tau \al-\xi)\, u_{t}]\,u.
\ea
\ee

Then the vector field $C=(C^{0},C^{1})$ is a conserved field for equation $(\ref{111})$.
\end{corollary}

{\bf Remarks:}

\rm 1. According to Theorem \ref{teo6}, given a projectable symmetry generator $X$ of equation (\ref{111}), from it we can construct an infinite number of nonprojectable symmetry generators, given by $X_{\lambda}=\lambda(u)X$, where $\lambda=\lambda(u)$ is a smooth function.

\rm 2. From Corollary \ref{cor2} it is easy to conclude that given a projectable symmetry generator $X$ of equation (\ref{111}), it is possible to obtain an infinite number of conservation laws, given by formulae (\ref{ncl}).

\subsection{Conservation laws for inviscid Burgers equation}\label{leis}

Here we shall use the Ibragimov's Theorem on conservation laws \cite{ib} to establish the conservation laws for equation (\ref{3}).

From Theorem \ref{teo2}, equation (\ref{3}) is self-adjoint and the Theorem \ref{teo5} can be employed in order to establish conservation laws for it.

From (\ref{ncl}) and (\ref{3}) a conserved vector is $C=(C^{0},C^{1})$, where
\bb\label{cl1}
\ba{l c l}
C^{0}&=&\left[\eta+(\tau a(u)-\xi)u_{x}\right]u,\\
\\
C^{1}&=&\left[\eta a(u)-(\tau a(u)-\xi)u_{t}\right]u.
\ea
\ee

With regard to the time and spatial translational invariance, it is easy to check that the conservation laws are trivial.
Let $A(u)$ be a function such that
\bb\label{2.5.1}
A'(u)=u\,a(u).
\ee
For the symmetry $X_{3}$, the conservation law is $D_{t}C^{0}+D_{x}C^{1}=0$, where:
$$
\ba{l c l}
C^{0}&=&[ta(u)-x]\,u\,u_{x},\\
\\
C^{1}&=&[x-ta(u)]\,u\,u_{t}.
\ea
$$

Equation (\ref{3}) is a first-order equation. Consequently, zero-order conservation laws are more important than first-order one. So we intend to simplify the components $C^{0}$ and $C^{1}$ in order to establish zero-order conservation laws for (\ref{3}).

Since $C^{0}=D_{x}(tA)-xD_{x}(u^{2}/2)$ and $C^{1}=D_{t}(xu^{2}/2)-tD_{t}(A)$, the conserved vector $C=(C^{0},C^{1})$ can be simplified using the fact
$$
\ba{lcl}
D_{t}C^{0}+D_{x}C^{1}&=&\ds{D_{x}(A)+tD_{t}D_{x}(A)-xD_{t}D_{x}\left(\f{u^{2}}{2}\right)}\\
\\
&&\ds{+D_{t}\left(\f{u^{2}}{2}\right)+xD_{x}D_{t}\left(\f{u^{2}}{2}\right)-tD_{x}D_{t}(A)}\\
\\
&=&\ds{D_{t}\left(\f{u^{2}}{2}\right)+D_{x}(A)}.
\ea
$$
It follows that
\bb\label{l1}
C^{0}=\f{u^{2}}{2},\,\,\,\,C^{1}=A,
\ee
where $A$ is given in (\ref{2.5.1}), provides a conserved vector for equation (\ref{3}).

Below we present, in a schematic form, the conservation laws associated to the Lie point symmetry generators $X_{4},\cdots,X_{8}$. First we present the conservation laws given by Theorem \ref{teo5}. In the following, we give the simplified vector employing the same procedure used in order to obtain the components (\ref{l1}).
\begin{enumerate}
\item  For the symmetry $X_{4}$, the components of the vector field given by Theorem \ref{teo5} are
$$
C^{0}=-\f{a(u)}{a'(u)}\,u+t\,a(u)\,u\,u_{x},\,\,\,\,C^{1}=-\f{a(u)^{2}}{a'(u)}\,u-t\,a(u)\,u\,u_{t}.
$$

The simplified components are
\bb\label{l2}
C^{0}=\f{a\,u}{a'},\,\,\,\,C^{1}=\f{a^{2}\,u}{a'}-A.
\ee

\item  For the symmetry $X_{5}$, the components of the vector field given by Theorem \ref{teo5} are
$$
C^{0}=\f{u}{a'(u)}-t\,u\,u_{x},\,\,\,\,C^{1}=\f{a(u)}{a'(u)}\,u+t\,u\,u_{t}.
$$

The simplified components are
\bb\label{l3}
C^{0}=\f{u}{a'},\,\,\,\,C^{1}=\f{a\,u}{a'}-\f{u^{2}}{2}.
\ee

\item  For the symmetry $X_{6}$, the components of the vector field given by Theorem \ref{teo5} are
$$
C^{0}=-\f{a(u)^{2}}{a'(u)}\,u+x\,a(u)\,u\,u_{x},\,\,\,\,C^{1}=-\f{a(u)^{3}}{a'(u)}\,u-x\,a(u)\,u\,u_{t}.
$$

The simplified components are
\bb\label{l4}
C^{0}=\f{a(u)^{2}}{a'(u)}\,u+A,\,\,\,\,C^{1}=\f{a(u)^{3}}{a'(u)}\,u.
\ee

\item  For the symmetry $X_{7}$, the components of the vector field given by Theorem \ref{teo5} are
$$
C^{0}=\f{x-t\,a(u)}{a'(u)}\,u+(t^{2}\,a(u)-t\,x)\,u\,u_{x},\,\,\,\,C^{1}=\f{x-t\,a(u)}{a'(u)}\,a(u)\,u-(t^{2}\,a(u)-t\,x)\,u\,u_{t}.
$$

The simplified components are
\bb\label{l5}
C^{0}=\f{x-ta(u)}{a'(u)}\,u+\f{t\,u^{2}}{2},\,\,\,\,C^{1}=\f{x-ta(u)}{a'(u)}\,a(u)\,u+2\,t\,A-\f{x\,u^{2}}{2}.
\ee

\item  For the symmetry $X_{8}$, the components of the vector field given by Theorem \ref{teo5} are
$$
C^{0}=\f{x-t\,a(u)}{a'(u)}\,a(u)\,u+(t\,x\,a(u)-x^{2})\,u\,u_{x},\,\,\,\,C^{1}=\f{x-t\,a(u)}{a'(u)}\,a(u)^{2}\,u-(t\,x\,a(u)-x^{2})\,u\,u_{t}.
$$

The simplified components are
\bb\label{l5}
C^{0}=\f{x-ta(u)}{a'(u)}\,a(u)\,u+x\,u^{2}-t\,A,\,\,\,\,C^{1}=\f{x-ta(u)}{a'(u)}\,a^{2}(u)\,u+x\,A.
\ee
\end{enumerate}

\section{Conclusion}

\

In this paper we considered the general problem on group classification of the general first order evolution equation (\ref{1.1}). We found the general classes of the quasi-self and self-adjoint equations of the type (\ref{1.1}). By using the recent Ibragimov's Theorem on conservation laws, we derive the general formulae to the conserved fields. Our main results are summarized in theorems \ref{teo1} - \ref{teo6}, corollaries \ref{cor1}, \ref{cor2} and in the conservation laws for inviscid Burgers equation established in section 3 (equations (\ref{l1}) - (\ref{l5})).

We believe that the research on group analysis of equations type (\ref{2.2.2}) can be promising. From Theorem 3, this equation is the most general first order evolution equation that admits self-adjoint equations. Following the Ibragimov's Theorem on conservation laws \cite{ib}, we have a closed form to express its conservation laws given by equation (\ref{cl1}).

From (\ref{2.2.2}) and the determining equations (\ref{deteq}) it is noted that we can obtain an underdetermined system of equations to be solved for 5 unknown functions $\al,\,\be,\,\tau,\,\xi$ and $\eta$. Thus, the set of symmetries is infinity.

With regard to the inviscid Burgers equation (\ref{3}), the set of the determining equations is also underdetermined, as we can observe in (\ref{3.2})-(\ref{3.3}). See also \cite{me1,ou}. From Theorem \ref{teo6} an infinite number of new symmetries of equation (\ref{3}) are presented supposing that $X_{\lambda}=\lambda(u)X$, where $\lambda(u)$ is a smooth function, and

$$X=\tau(t,x)\f{\p}{\p t}+\xi(t,x)\f{\p}{\p x}+\eta(t,x,u)\f{\p}{\p u}$$
is a projectable symmetry generator of (\ref{3}). This just is one more possible ansatz to determine Lie point symmetries of (\ref{3}). From Corollary \ref{cor2}, the corresponding conservation laws associated to the symmetries given by Theorem \ref{teo6} are established.

A natural question that arises is: which more ansatz can someone use in order to obtain more general symmetries of the inviscid Burgers equation? This is a question that, hopefully, can inspirit some more progress in this area.

\section*{Acknowledgments}

\

I would like to thank the referee of this paper. The referee not only read this article very carefully, but also contributed with numerous comments/corrections that improved the paper.

I am grateful to Priscila L. da Silva for her careful reading of this paper.

\end{document}